\documentstyle[12pt,epsf,rotate]{article}
\begin{document}
\pagestyle{empty}

\hfill{\sf Hot Points in Astrophysics}

\hfill{\sf JINR, Dubna, Russia, August 22-26, 2000}

\vspace*{2.cm}

\centerline{\bf Isolated Neutron Stars in the Galaxy}

\vspace*{0.2cm}

\begin{center}

Sergei B. Popov$^{\dagger}$, Mikhail E. Prokhorov$^{\dagger}$,
 Monica Colpi$^{\ddagger}$, Aldo Treves$^{*}$, Roberto Turolla$^{**}$
and Vladimir M. Lipunov$^{\dagger}$\\[0.3cm]

{\it $^\dagger$Sternberg Astronomical Institute, Universitetski pr. 13, 
119899, Moscow, Russia \\
$^{\ddagger}$University Milano-Bicocca, Milan, Italy\\
$^{*}$Universit\'a  dell'Insubria, 22100, Como, Italy\\
$^{**}$University of Padova, Via Marzolo 8, 35131, Padova, Italy}

\end{center}

\vspace*{0.2cm}

\begin{abstract}

 In this article we briefly review our recent results on evolution 
and properties of isolated
neutron stars (INSs) in the Galaxy.

As the first step we calculate a {\it census} of INSs in our Galaxy. 
 We infer a lower bound for the mean  kick velocity of NSs,
$ <V> \sim $(200-300)  ${\rm km\,s^{-1}}$.
The same conclusion is reached for both a constant magnetic field
($B\sim 10^{12}$ G) and for a magnetic field decaying exponentially with a
timescale $\sim 10^9$ yr. These results,
moreover, constrain the fraction of low velocity stars, which
could have escaped pulsar statistics, to $\sim$few percents.

Then we show that
the range of minimum  value of magnetic moment, $\mu_b$:
$\sim 10^{29.5}\ge \mu_b \ge 10^{28} \, {\rm G}\, {\rm cm}^3$, 
and the characteristic  decay time, $t_d$:
$\sim 10^8\ge  t_d \ge 10^7\, {\rm yrs}$, can be excluded
assuming the standard
initial magnetic momentum, $\mu_0=10^{30} \, {\rm G}\, {\rm cm}^3$,
if accreting INSs are observed.
For these parameters
an INS would never reach the stage of accretion from the interstellar
medium even for a low space velocity of the star
and high density of the ambient plasma. 
The range of excluded parameters increases for lower values of $\mu_0$.

It is shown that old accreting INSs  become more abundant than young cooling
INSs at X-ray fluxes below $\sim 10^{-13}$ erg cm$^{-2}$ s$^{-1}$.
We can predict that about one accreting INS per square
degree should be observed at the {\it Chandra} and {\it Newton} flux limits
of  $\sim 10^{-16}$ erg cm$^{-2}$ s$^{-1}.$
The weak {\it ROSAT} sources, associated with INSs,
can be young cooling objects, if the NSs birth rate
in the solar vicinity during the last $\sim 10^6$ yr was much higher than
inferred from radiopulsar observations.

\end{abstract}

\vspace*{0.5cm}

\section*{Introduction}

Despite intensive observational campaigns, no irrefutable identification 
of an isolated
accreting neutron star (IANS) has been presented so far. 
Although six soft, weak sources, 
which are associated with isolated NSs, have been found in
ROSAT fields, present X-ray and optical data do not allow an unambiguous
determination of the physical mechanism responsible for their emission. 
These sources
can be powered either by accretion of the interstellar gas onto old 
($\approx 10^{10}$ yr) 
INSs or by the release of internal energy in relatively young ($\approx
10^6$ yr) cooling INSs (see \cite{t2000} for recent review). 
ROSAT candidates, 
although relatively bright (up to $\approx 1 \ {\rm cts\,s}^{-1}$), 
are intrinsically dim 
and their inferred luminosity ($L\approx 10^{31} \ {\rm erg\, s}^{-1}$)
is close to that expected from both close-by cooling and (most luminous)
accreting INSs. 
Up to now only two optical counterparts have been possibly identified 
(RX J 1856-3754,  \cite{wm97}; 
RX J 0720-3125,  \cite{kk98}) and in both cases an
optical excess over the low-frequency tail of the black body X-ray spectrum 
has been reported.
While detailed multiwavelength observations with next-generation instruments 
may indeed be
the key for assessing the true nature of these sources, other, indirect, 
approaches may be
used to discriminate in favor of one of the two scenarios proposed so far.

Since early 90$^s$, when in \cite{tc91} it was suggested to search for IANSs
with ROSAT satellite, several investigations on INSs have been done (see
for exapmle \cite{mb94}, \cite{mann96}, and \cite{t2000} for a review). 
Here we present our recent results in that field.

\section*{Neutron Star Census}

We have investigated, \cite{p2000}, how the present distribution of
NSs in the different stages (Ejector, Propeller,
Accretor and Georotator, see \cite{l92}) depends on the star mean velocity at
birth (see Fig.~1). The fraction of
Accretors  was used to estimate the number of sources within 140
pc from the Sun which should have been detected by ROSAT. Most
recent analysis of ROSAT data indicate that no more than $\sim 10$
non--optically identified sources can be accreting old INSs. This
implies that the 
average velocity of the INSs population at birth has to
exceed $\sim 200 \ {\rm km\, s^{-1}}$, a figure which is
consistent with those derived from radio pulsars statistics. We
have found that this lower limit on the mean kick velocity is
substantially the same either for a constant or a decaying
$B$--field, unless the decay timescale is shorter than $\sim 10^9$
yr. Since observable accretion--powered INSs are slow objects, our
results exclude also the possibility that the present velocity
distribution of NSs is richer in low--velocity objects with
respect to a Maxwellian. The paucity of accreting INSs seem
to lend further support in favor of NSs as
fast objects.

\begin{figure}[h!]
\epsfxsize=10cm
\centerline{\rotate[r]{\epsfbox{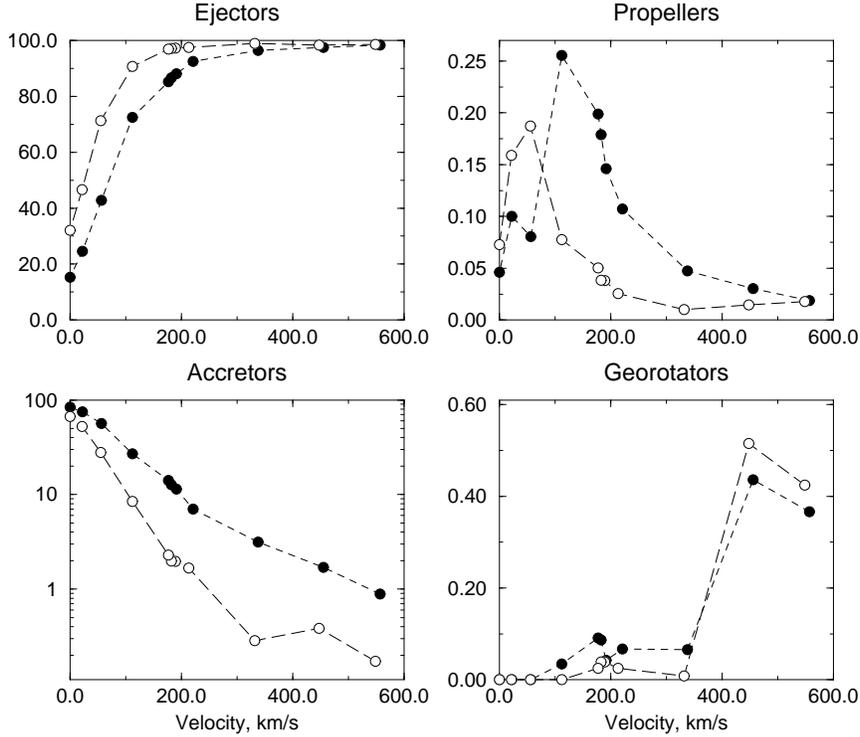}}}
\caption{Fractions of NSs (in percents) 
on different stages vs. the mean kick
velocity for $\mu_{30}=0.5$ (open circles) and $\mu_{30}=1$
(filled circles); typical statistical uncertainty for ejectors and
accretors is $\sim $ 1-2\%. Figures are plotted for constant magnetic field.}
\end{figure}

\section*{Magnetic Field Decay}

Magnetic field decay can operate in INSs. Probably, some
of observed ROSAT INS candidates represent such exapmples (\cite{kp97},
\cite{w97}) 
 We tried to evaluate the region of parameters which is
excluded for models of the exponential
magnetic field decay in INSs using
the possibility that some of  ROSAT 
soft X-ray sources are indeed old AINSs.  

In this section  we follow the article \cite{pp2000}.

\begin{figure}[h!]
\epsfxsize=10cm
\centerline{\rotate[r]{\epsfbox{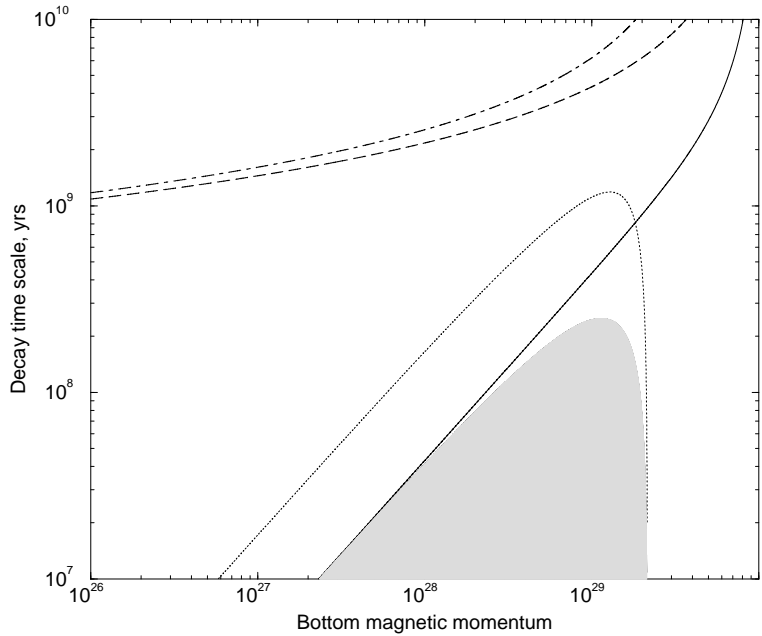}}}
\caption{ The characteristic time scale of the magnetic field decay, $t_d$, vs.
bottom magnetic moment, $\mu_b$.
In the hatched region Ejector life time, $t_E$, 
is greater than $10^{10} {\rm yrs}$.
The dashed line corresponds to $t_H=t_d\cdot \ln \left( \mu_0/\mu_b
\right)$, where $t_H=10^{10}$ years. The solid line corresponds to
$p_E(\mu_b)=p(t=t_{cr})$, where $t_{cr}=t_d\cdot \ln \left(
\mu_0/\mu_b \right)$. Both the lines and hatched region
are plotted for $\mu_0=10^{30} {\rm G} \, {\rm cm}^{-3}$. 
The dash-dotted line is the same as the dashed one, 
but for $\mu_0=5\cdot 10^{29} 
\, {\rm G} \, {\rm cm}^3$.
The dotted line shows the border of the ``forbidden'' region for $\mu_0=5\cdot
10^{29}  \, {\rm G} \, {\rm cm}^3$. See details in \cite{pp2000}.}
\end{figure} 

Here the field decay is assumed to have an exponential shape:

\begin{equation}
\mu=\mu_0\cdot e^{-t/t_d}, \, {\rm for} \, \mu > \mu_b
\label{eq:mu(t)}
\end{equation} 
where $\mu_0$ is the initial magnetic moment 
($\mu=\frac12 B_p R_{NS}^3$, here $B_p$ is the polar magnetic field,
$R_{NS}$ is the NS radius), $t_d$ is the characteristic time
scale of the decay, and $\mu_b$ is the bottom value of the 
magnetic momentum which is reached at the time 

\begin{equation}
t_{cr}=t_d\cdot \ln\left( \frac{\mu_0}{\mu_b} \right)
\end{equation}
and does not change after that. 

The intermediate values of $t_d$ ($\sim 10^7-10^8 \, {\rm yrs}$)
in combination with the intermediate values of
$\mu_b$ ($\sim 10^{28}-10^{29.5} \, {\rm G} \, {\rm cm}^3$)
for $\mu_0=10^{30} \, {\rm G}\, {\rm cm}^3$
can be excluded for progenitors of isolated accreting NSs
because NSs with such parameters would always remain
on the Ejector stage and never pass to the accretion stage
(see Fig.~2).
Even if all modern candidates are not accreting objects,
the possibility of limitations of magnetic field
decay models based on future observations of isolated accreting NSs
should be addressed. 

 For higher $\mu_0$ NSs should reach the stage of Propeller (i.e. $p=p_E$,
where $p_E$-- is the Ejector period) even
for $t_d < 10^8 $ yrs, for weaker fields 
the ``forbidden'' region becomes
wider. Critical period, $p_E$, corresponds to transision from the Propeller
stage to the stage of Ejector, and is about 10-25 seconds for typical
parameters.
The results are dependent on the initial magnetic field, $\mu_0$, 
the ISM density, $n$, and NSs velocity, $V$. So here different ideas can be
investigated. 

In fact the limits obtained above are even stronger than
they could be in nature, i.e. ``forbidden'' regions can be wider, 
because we did not take into account
that NSs can spend 
some significant time (in the case with field decay)
at the propeller stage (the spin-down rate at this stage is very
uncertain, see the list of formulae, for example, in \cite{lp95}
or \cite{l92}). 
The calculations of this effect for
different models of non-exponential field decay were studied separately
in \cite{pp2001}.

Note that there is another reason due to which
a very fast decay down to small values of $\mu_b$ can also be
excluded, because this would lead to a huge amount of accreting isolated
NSs in drastic contrast with observations. This situation
is similar to the ``turn-off'' of the magnetic field of an INS
(i.e., quenching any magnetospheric effect on the accreting matter). So
for any velocity and density distributions we should expect 
significantly more accreting isolated NSs than we know from ROSAT
observations
(of course, for high velocities X-ray sources will be very dim, but close
NSs can be observed even for velocities $\sim 100$ km s$^{-1}$).

\section*{Log N -- Log S distribution}

In this section  we briefly present our new results on INSs, \cite{p2001}.

We compute and compare the $\log N$ -- $\log S$ distribution of both
accreting and cooling NSs, to establish the relative
contribution of the two populations to the observed number counts.
Previous studies derived the $\log N$ -- $\log S$
distribution of accretors (\cite{tc91}; \cite{mb94};
\cite{mann96}) assuming  a NSs velocity distribution rich in slow
stars ($v< 100 \ {\rm km\, s}^{-1}$). More recent measurements of pulsar
velocities (e.g. \cite{ll94})
and upper limits on the observed number of accretors in ROSAT surveys 
point, however, to a larger NS mean velocity
(see \cite{t2000} for a critical discussion).
Recently Ne\"uhauser \& Tr\"umper (\cite{nt99})
compared the number count distribution of the
ROSAT isolated NS  candidates  with those of accretors and coolers.
In \cite{p2001} we address these issues in  greater detail, also
in the light of the latest contributions to the modeling of the evolution
of Galactic NSs.

Our main results for AINSs are presented in Fig.~3.

\begin{figure}[h!]
\epsfxsize=10cm
\centerline{\rotate[r]{\epsfbox{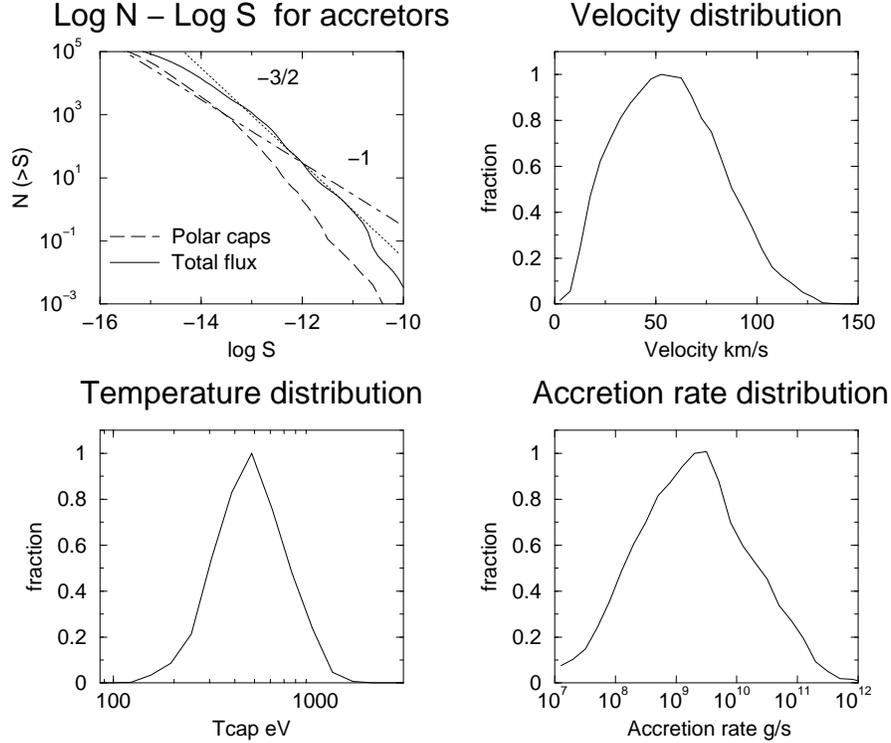}}}
\caption{Upper left panel: the $\log N$ -- $\log S$ distribution for
accretors within 5 kpc from the Sun. The two curves refer to total emission
from the entire star surface and
to polar cap emission in the range 0.5-2 keV; two straight lines with slopes -1 and -3/2
are also shown for comparison. From top right to bottom right: the velocity,
effective temperature and accretion rate distributions of accretors; all
distributions are normalized to their maximum value.}
\end{figure}

Using  ``standard'' assumptions
on the velocity, spin period and magnetic field parameters,
the accretion scenario can not explain the observed properties of the six
ROSAT candidates.

A key result of our  statistical analysis is that
accretors should  eventually become more abundant than coolers
at fluxes below
$10^{-13}$ erg cm$^{-2}$ s$^{-1}$.

\section*{Conclusions}

 INSs are now a real {\it Hot Point in Astrophysics}.
We tried to show how these objects are related with models
of magnetic field decay, and with recent anf future X-ray observations.

 Observed candidates propose ``non-standard'' properties of 
NSs. And future observations with XMM (Newton) and Chandra satellites
can give more important facts.

\section*{Acknowledgments}
We wish to thank  J. Lattimer, E. van den Heuvel and
S. Campana for useful discussions. 
SBP and MEP also thank the University of Insubria for
financial support and the Universities of  Milano-Bicocca and  Padova
together with the Brera Observatory (Merate) for their kind hospitality.
The work of SBP, VML and MEP was supported through grant RFBR 00-02-17164
and NTP Astronomy grants 1.4.4.1. and 1.4.2.3.

\end{document}